# A Parallel Bitstream Generator for Stochastic Computing

Yawen Zhang, Runsheng Wang, Xinyue Zhang, Zherui Zhang, Jiahao Song,
Zuodong Zhang, Yuan Wang, and Ru Huang
Key Laboratory of Microelectronic Devices and Circuits (MOE),
Institute of Microelectronics, Peking University, Beijing 100871, China
Email: r.wang@pku.edu.cn; wangyuan@pku.edu.cn

*Abstract* — **Stochastic computing (SC) presents high error tolerance and low hardware cost, and has great potential in applications such as neural networks and image processing. However, the bitstream generator, which converts a binary number to bitstreams, occupies a large area and energy consumption, thus weakening the superiority of SC. In this paper, we propose a novel technique for generating bitstreams in parallel, which needs only one clock for conversion and significantly reduces the hardware cost. Synthesis results demonstrate that the proposed parallel bitstream generator improves 2.5× area and 712× energy consumption.**

## I. Introduction

As a promising alternative to conventional binary computing, stochastic computing (SC) [1-2] which represents the data by the probability of a "1" in bitstreams, has high error tolerance under low operating voltage [3] showing great potential applications in the internet-of-things (IoT). What is more, SC achieves complex arithmetic operations with simple logic gates. For example, multiplication can be realized by AND gates and scaled addition can be realized by MUX gates, as shown in **Fig. 1**. However, the bitstream generator occupies over 80% area [4-5] and large energy consumption of whole SC circuits, which reduces the advantage of SC. Therefore, reducing the area and energy consumption of bitstream generators is one of the key issues in stochastic computing.

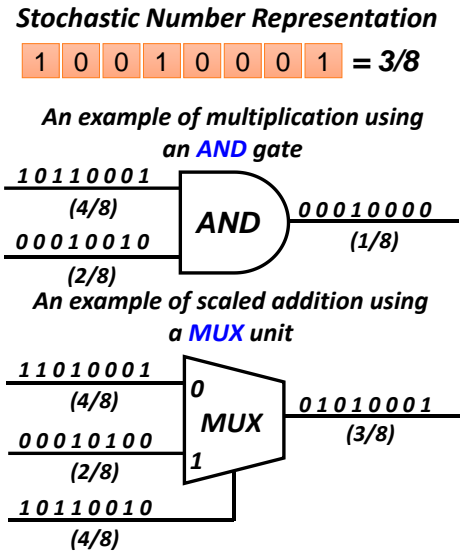

Fig. 1. Stochastic number representation. Examples of multiplication and addition in stochastic computing.

In this paper, we address these problems through a novel parallel bitstreams generator based on thermometer coding, which only requires a simple decoder to synchronously decode a binary number into bitstreams. Compared with the traditional bitstream generator, the proposed parallel bitstream generator achieves low hardware cost and low energy consumptions.

## II. Accuracy of Different Coding Methods

Traditional approaches convert binary numbers to stochastic bitstreams including a pseudo-random number source and a comparator. Pseudo-random number source has three coding methods (**Fig. 2**), which are linear feedback shift register (LFSR) sequences [1], low-discrepancy (LD) sequences [6] and thermometer coding method [7].

| | Generator | Two independent bit streams | | | | | | | | | | | | | | | |
|---|---|---|---|---|---|---|---|---|---|---|---|---|---|---|---|---|---|
| LFSR seq. | LFSR (seed1) | 0 | 15 | 11 | 9 | 8 | 4 | 2 | 1 | 12 | 6 | 3 | 13 | 10 | 5 | 14 | 7 |
| | LFSR (seed2) | 12 | 6 | 3 | 13 | 10 | 5 | 14 | 7 | 0 | 15 | 11 | 9 | 8 | 4 | 2 | 1 |
| LD seq. | Counter | 0 | 1 | 2 | 3 | 4 | 5 | 6 | 7 | 8 | 9 | 10 | 11 | 12 | 13 | 14 | 15 |
| | Counter+Conv. | 0 | 8 | 4 | 12 | 2 | 10 | 6 | 14 | 1 | 9 | 5 | 13 | 3 | 11 | 7 | 15 |
| Thermometer coding | Counter | 0 | 1 | 2 | 3 | 0 | 1 | 2 | 3 | 0 | 1 | 2 | 3 | 0 | 1 | 2 | 3 |
| | Counter | 0 | 0 | 0 | 0 | 1 | 1 | 1 | 1 | 2 | 2 | 2 | 2 | 3 | 3 | 3 | 3 |

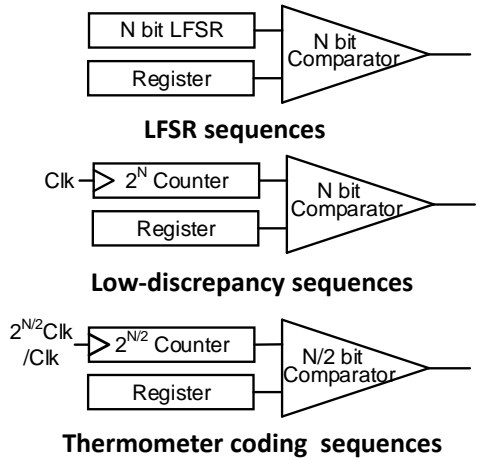
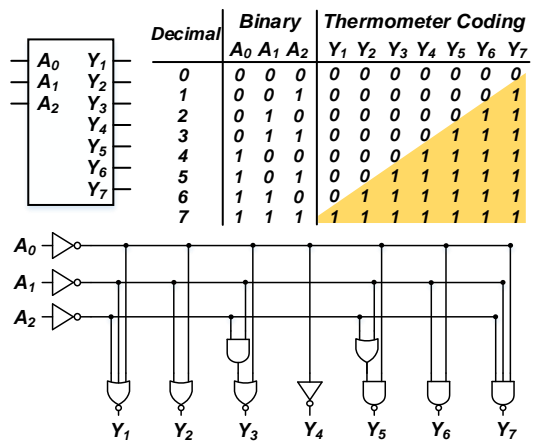

Fig. 2. Three coding methods in stochastic computing. Traditional bitstream generators with three coding methods and the proposed parallel bitstream generator with thermometer coding.

Ref. [6] shows that the accuracy of bitstreams based on LD sequences is better than that based on LFSR sequences, due to their low correlation. However, as shown in **Fig. 3**, LFSR sequences actually can also achieve relatively low discrepancy and low correlation by carefully selecting different initial seeds. The mean square error (MSE) of two input multiplication results (AND gate calculation results, $p_{x_1} \Lambda p_{x_2}$) is used to evaluate the correlation of two bitstream generators. Taking the 8-bit LFSR sequences as an example, as shown in **Fig. 4**, the MSE with different LFSR initial seeds is quite different. For the min MSE case, the error of most multiplication results is almost zero, which means that the correlation between these two LFSR sequences is very low. For a fair comparison, in the following, all evaluations are based on the min MSE LFSR sequences.

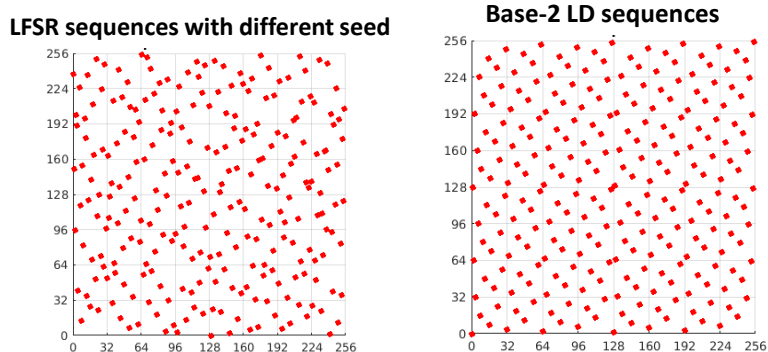

Fig. 3. Discrepancy of LFSR sequences and base-2 LD sequences.

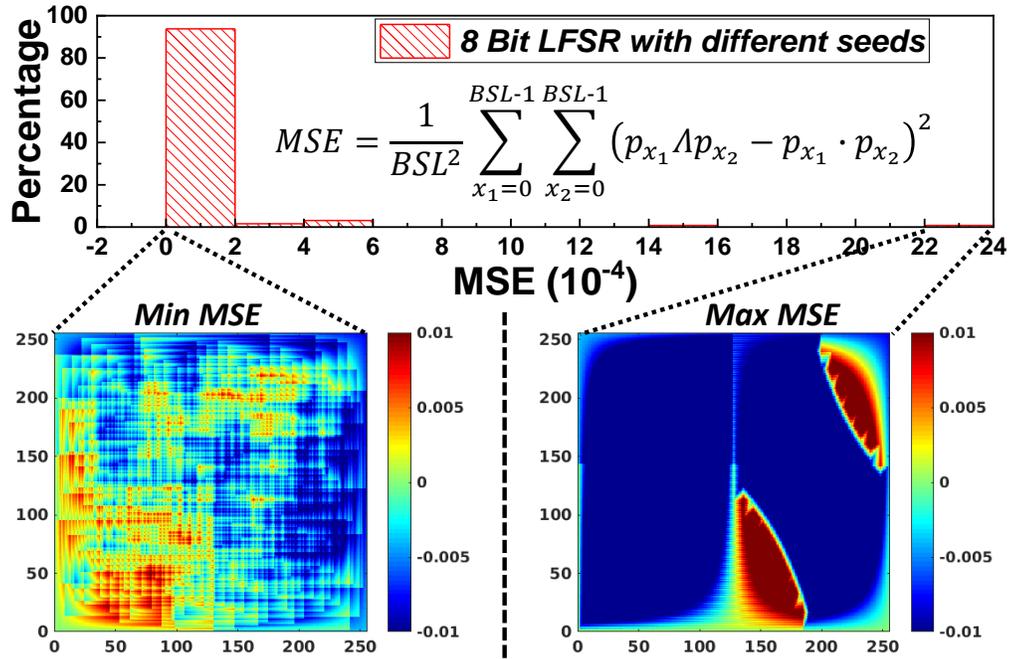

Fig. 4. The MSE calculation results using different 8-bit LFSR initial seeds. The calculation error of AND gate ($p_{x_1} \Lambda p_{x_2}$) using LFSR sequences with the min (left) and the max (right) MSE.

**Fig. 5** shows the MSE of two 4-bit input multiplication based on three coding methods. It is found that the MSE of LFSR sequences is the smallest in 16 bitstream length (BSL). With the BSL increasing, the MSE of LD sequences decreases rapidly and becomes smaller than that of LFSR sequences. Due to the rounding error, the MSE of thermometer coding is large, but it instantaneously decreases to 0 when BSL reaches 256. This also illustrates that thermometer coding does not have progressive accuracy [6]. The error of two 4-bit input multiplication under different BSL displays in **Fig. 6**.

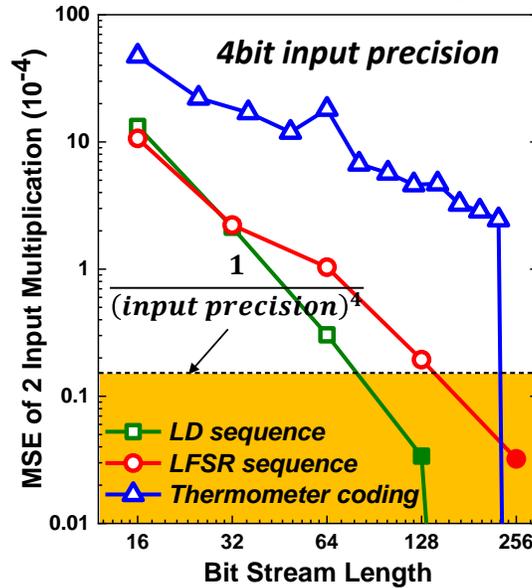

Fig. 5. The MSE of two 4bit input multiplication results in three coding methods.

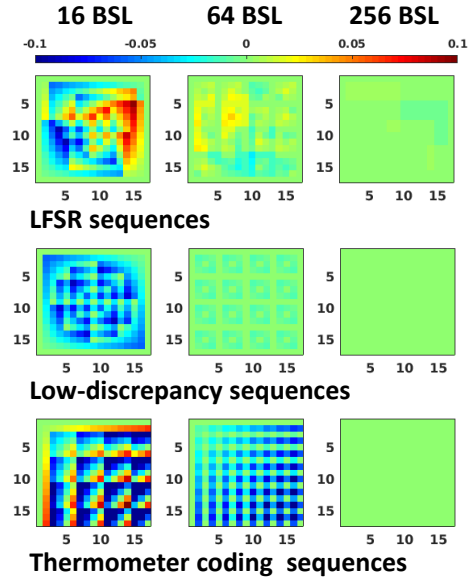

Fig. 6. The calculation error of multiplication results with different BSL in three coding methods.

### III. PARALLEL BITSTREAM GENERATOR

By observing three existing coding methods [1][6-7] of bitstreams in SC, we found that, to some extent, they are actually all deterministic in practice. Especially, the thermometer coding method is the most regular and achieves complete accuracy after reaching a certain BSL. Since the thermometer coding method is simple and regular, here we propose a parallel bitstream generator based on thermometer coding method, which uses a binary-to-thermometer decoder as shown in **Fig. 2**. The proposed parallel bitstream generator reduces the latency to only one clock and achieves less hardware cost as well.

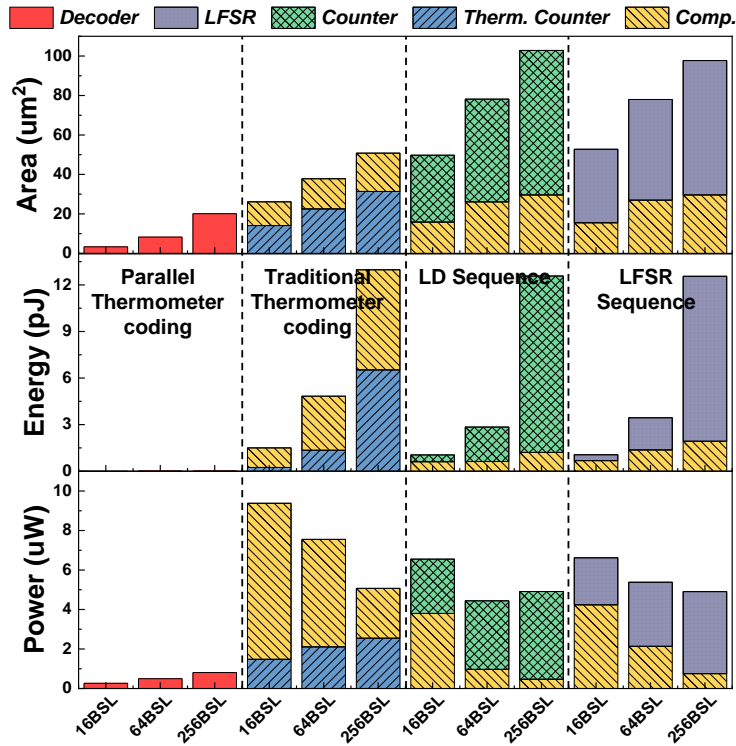

Fig. 7. Synthesis results in different bitstream generators, in terms of area and energy.

We use Synopsys Design Compiler to synthesize different bitstream generators based on TSMC 40nm technology library. **Fig. 7** shows the synthesis results of bitstream generators in terms of area, power and energy. Compared with the traditional bitstream generator, the proposed parallel bitstream generator greatly reduces area and energy consumption. Though traditional bitstream generators can share the same pseudo-random number source, the comparators cannot be shared [8], and thus their hardware costs are still higher than parallel bitstream generator. Due to the simplicity of decoder and full parallelism, the proposed parallel bitstream generator performs much better than traditional bitstream generators in terms of area (**Fig. 8**), power (**Fig. 9**), and energy (**Fig. 10**).

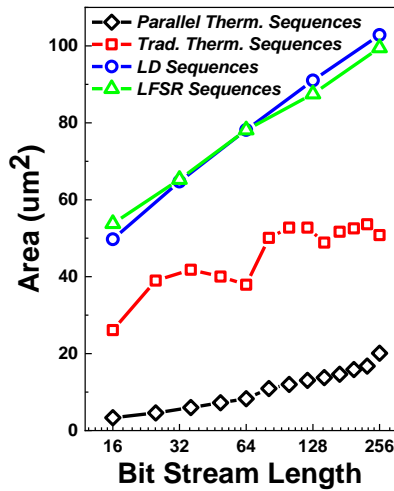
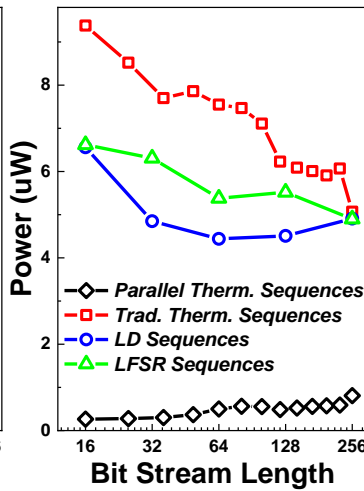
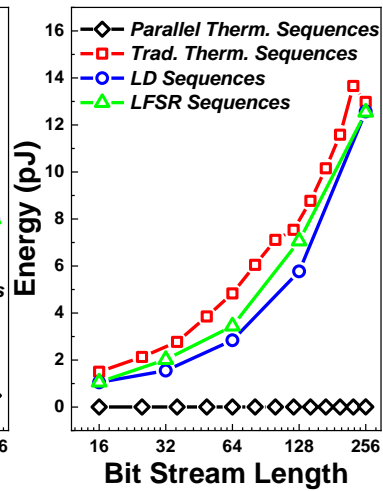

Fig. 8. Area synthesis results of bitstream generators based on different coding methods.

Fig. 9. Power synthesis results of bitstream generators based on different coding methods.

Fig. 10. Energy synthesis results of bitstream generators based on different coding methods.

In order to fairly compare the synthesis results based on different coding methods, here we assume that the MSE of two 4-bit input multiplication cannot exceed $1/(input\ precision)^4$, as shown in **Fig. 5**. With this limited accuracy, as shown in **TABLE**, the proposed parallel bitstream generator achieves at least 2.5× area improvements, 5.6× power improvements, and 712× energy improvements compared with traditional bitstream generators. It is worth noting that the bit generation efficiency of the proposed parallel bitstream generator reaches 31605 bit/pJ in 256 BSL, which achieves at least 1425× improvements.

| Coding Method | Component | Bit Stream Length | Bit Precision | Power ($\mu$W) | Area ($\mu m^2$) | Latency ($\mu s$) | Energy (pJ) | Bit Generation Efficiency (bit/pJ) |
|---|---|---|---|---|---|---|---|---|
| LFSR Sequences | 4bit LFSR+Comp. | 16 | 16 | 6.62 | 53.8 | 0.16 | 1.059 | 15.11 |
|  | 6bit LFSR+Comp. | 64 | 64 | 5.38 | 78.15 | 0.64 | 3.443 | 18.59 |
|  | 8bit LFSR+Comp. | 256 | 256 | 4.9 | 99.49 | 2.56 | 12.54 | 20.41 |
| LD Sequences | 4bit Counter+Comp. | 16 | 16 | 6.56 | 49.74 | 0.16 | 1.05 | 15.24 |
|  | 6bit Counter+Comp. | 64 | 64 | 4.44 | 78.15 | 0.64 | 2.842 | 22.52 |
|  | 7bit Counter+Comp. | 128 | 128 | 4.51 | 91.02 | 1.28 | 5.773 | 22.17 |
|  | 8bit Counter+Comp. | 256 | 256 | 4.91 | 102.8 | 2.56 | 12.57 | 20.37 |
| Traditional Thermometer Coding | 2Bit Counter+Comp. | 16 | 4 | 9.38 | 26.11 | 0.16 | 1.501 | 10.66 |
|  | 3Bit Counter+Comp. | 64 | 8 | 7.55 | 37.93 | 0.64 | 4.832 | 13.25 |
|  | 4Bit Counter+Comp. | 256 | 16 | 5.07 | 50.8 | 2.56 | 12.98 | 19.72 |
| **Parallel Thermometer Coding (This work)** | 3-4 Decoder | 16 | 4 | **0.265** | 3.352 | 0.01 | 0.003 | 6038 |
|  | 4-8 Decoder | 64 | 8 | **0.5** | 8.291 | 0.01 | 0.005 | 12800 |
|  | 5-16 Decoder | 256 | 16 | **0.81** | 20.11 | 0.01 | 0.008 | 31605 |

IV. SUMMARY

In this paper, a parallel bitstream generator using a binary-to-thermometer coding decoder is proposed for the first time. We also provide a reasonable analysis method for the accuracy of different coding methods in SC. Comparing different bitstream generators with limited accuracy, the proposed parallel bitstream generator achieves at least 2.5× area improvements and 712× energy improvements.

ACKNOWLEDGMENTS

This work was partly supported by NSFC (61522402 and 61421005) and the 111 Project (B18001). The authors would like to thank Weikang Qian for the helpful discussions.

REFERENCES

[1] J. P. Hayes, DAC, 2015; [2] W. Qian, et al., Trans. on Computers, 2011; [3] Y. Zhang, et al., IEDM, 2017; [4] P. Li, et al., VLSI Systems, 2014; [5] M. H. Najafi, et al., VLSI Systems, 2017; [6] A. Alaghi and J. P. Hayes, DATE, 2014; [7] D. Jens on and M. Riedel, ICCAD, 2016; [8] M. Yang, et al., ISVLSI, 2018.